\documentclass[prb,showpacs,twocolumn,amsmath]{revtex4}
\usepackage{latexsym}
\usepackage{graphicx}

\newcommand{\be}{\begin{equation}}

\newcommand{\ee}{\end{equation}}

\newcommand{\reff}[1]{(\ref{#1})}

\newcommand{\giro}[1]{\stackrel{\circ}{#1}}

\begin{document}

\title{The $D_2$ point group of the Two-Rotors Model and  scissors modes of negative parity}

\author{ Fabrizio Palumbo}

\affiliation{
INFN Laboratori Nazionali di Frascati, 00044 Frascati, Italy}

\begin{abstract}

The intrinsic Hamiltonian of the Two-Rotors Model as a quantized classical system  in which the rotors are abstract rigid bodies has a  $D_2$ point symmetry and then its eigenstates come in quadruplets. Only one member of each of the two lowest quadruplets was investigated so far, the ground state and the scissors mode. I determine the whole quadruplets which turn out to  contain states of negative parity. 

When the rotors are made of particles, the $D_2$ symmetry is  broken. {\it The actual existence of the new states of the multiplets in atomic nuclei depends on the existence of  excited states of the neutron and proton fluids separately odd under inversion of the intrinsic coordinates}. The energies of the new states are $\omega+e, \omega +2e$ when one or both rotors have negative parity respectively, where $\omega$ is the scissors excitation energy and $e$ is the excitation energy of a single rotor with negative parity. Nonvanishing transitions occur only between states whose energies differ by $\omega$.
\end{abstract}

\maketitle

\section{ Introduction}

The Two-Rotors Model (TRM)  describes the dynamics of two rigid bodies rotating with respect to each other under an attractive force around their centers  of mass. It was devised as a  model for deformed atomic nuclei, in which case  the rigid bodies represent the proton and neutron systems~[\onlinecite{LoIu}].  The low lying excited  states predicted by  this model [\onlinecite{Hilt}] were observed  in all deformed atomic nuclei~[\onlinecite{Bohle}] and  were called by B. Barrett scissors modes, see Fig.1.

   \begin{figure} 
  \begin{center}
    \begin{tabular}{cc}
    \includegraphics[width=3cm]{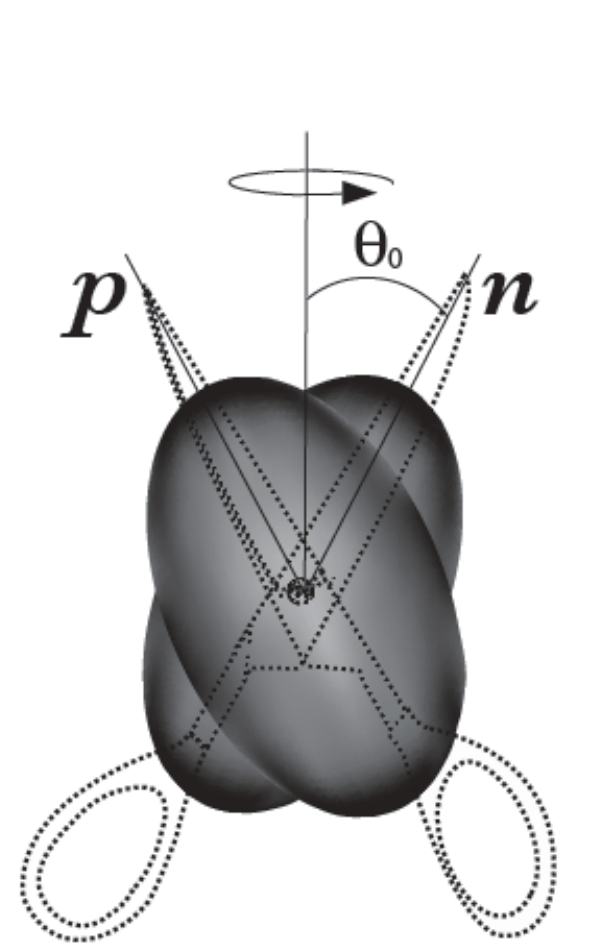}
    \end{tabular}
  \end{center}
 \caption{Scissors modes in the Two-Rotors Model:  the proton (p) and neutron (n) rotors  precess around the bisector  of their axes. }
\end{figure}

By analogy similar collective excitations were predicted in several other systems~[\onlinecite{Guer}] and clearly observed in Bose-Einstein condensates~[\onlinecite{Mara}] and very recently in dipolar quantum droplets of Dysprosium atoms~[\onlinecite{Ferr}]. Moreover an application of the TRM to the evaluation of the magnetic susceptibility of single domain magnetic nanoparticles stuck in rigid matrices has given results  compatible with a vast body of experimental data with an agreement  in some cases surprisingly good~[\onlinecite{Hata}].

Later   it was found that, according to the Brink hypotesis~[\onlinecite{Brin}], scissors modes live also on excited nuclear levels~[\onlinecite{Krti, Adek}]. This finding might be relevant to get experimental information on the states of the TRM investigated in the present paper. 

Another indication of the stability of scissors modes comes from the  recent confirmation~[\onlinecite{Beck}] of the existence of the $J=2$ member  of its rotational band, already predicted in~[\onlinecite{LoIu}]. Most important for the present work would be observation of the $J=3$ member that can  tell whether scissors modes are entangled as predicted~[\onlinecite{Palu}]  by the TRM.   Indeed  Fig.1,  while very suggestive, does not give a complete  representation  of the TRM states, because  the TRM Hamiltonian has a double well potential and then at the classical level two states localized at  the two minima which give rise at the quantum level to the entanglement shown in Fig.2.

Higher lying intrinsic states have a nontrivial  intrinsic structure that, unlike that of most collective models, cannot be described in terms of many phonons. For this reason they present a theoretical interest irrespective of their possible relevance to phenomenology. 

At this point I must notice that  the observed B(M1) strength of scissors modes shows  a highly  complex structure [\onlinecite{Balb0}]. On the theoretical side soon after their discovery it was suggested that their resonance should be split in the presence of triaxial deformation [\onlinecite{Palu0}]. Afterwards new orbital collective states related to scissors modes were introduced and called "twists" [\onlinecite{Romp}] and  spin collective states called spin scissors modes [\onlinecite{Otsu}]. The above is only a short sample of a huge literature on the subject whose review is out of the scope of the present work. The conclusion is that while  considerable progress has been made in several cases, especially showing how clusters of levels can be generated, a unified description of the fine structure of scissors modes is still lacking. It would presumably require the simultaneous  inclusion of all the mechanisms analized in the above papers and their interplay~[\onlinecite{Palumbo}]. 

While the TRM can give results which for several observables can be only  qualitative, within such a limitation it can be a  useful guide in the construction of a microscopic theory.  Moreover   I deem it interesting to explore  the limits of validity of a model after part of its spectrum has been experimentally confirmed even when the realisation of the other part might appear to be highly speculative. 

I therefore come  back to the TRM and I first consider the case in which the rotors are abstract bodies, without regard to their realization in terms of particles, and I will later discuss its application to atomic nuclei. I restrict myself to  rotors  which both have axial symmetry and  are invariant under inversion of their  symmetry axes, or equivalently under a  rotation through $\pi$ about an axis perpendicular to the symmetry axis. Even when the rotors have such a symmetry, however,  the   two rotors system does not have it. It does not even have triaxial symmetry,  because a rotation through $\pi$ about the bisector of the rotor axes of the TRM  interchanges protons and neutrons.  
 
 The intrinsic Hamiltonian of a triaxial system  is invariant under rotations through $\pi$ about each of its  principal axes. These operations, ${\mathcal R}_1(\pi), {\mathcal R}_2(\pi), {\mathcal R}_1(\pi) {\mathcal R}_2(\pi)={\mathcal R}_3(\pi)$ together with the identity are a realization of the  $D_2$ point group~[\onlinecite{Bohr}]. The intrinsic Hamiltonian of the TRM is instead  invariant under the  separate and the simultaneous inversion of the  rotors axes, and also these operations together with the identity are a realization of the $D_2$ point group. Because of such a symmetry  the eigenstates  come in quadruplets, but until now only one member of the 2 lowest quadruplets has been studied, the ground state and the scissors mode. In the present paper I determine completely the 2 lowest quadruplets. Each of them turns out to consist  of two doublets  of opposite parity.  

  This finding requires a preliminary discussion. The inversion  symmetry in previous papers was imposed~[\onlinecite{LoIu}] by requiring that for each rotor ${\mathcal R}(\pi)\psi= \psi$,  where $\psi$ is the wave function of the rotor and ${\mathcal R}(\pi)$ is an operator  which performs a rotation through $\pi$ about an axis perpendicular to its symmetry axis. The point is that such a condition is too restrictive, because  it is sufficient that  the probability of observing  one direction of a rotor axis does not depend on its orientation, namely that ${\mathcal R}(\pi)\psi = \exp(i\alpha) \psi$ for an arbitrary phase $\alpha$.

      \begin{figure} 
  \begin{center}
    \begin{tabular}{cc}
    \includegraphics[width=8cm]{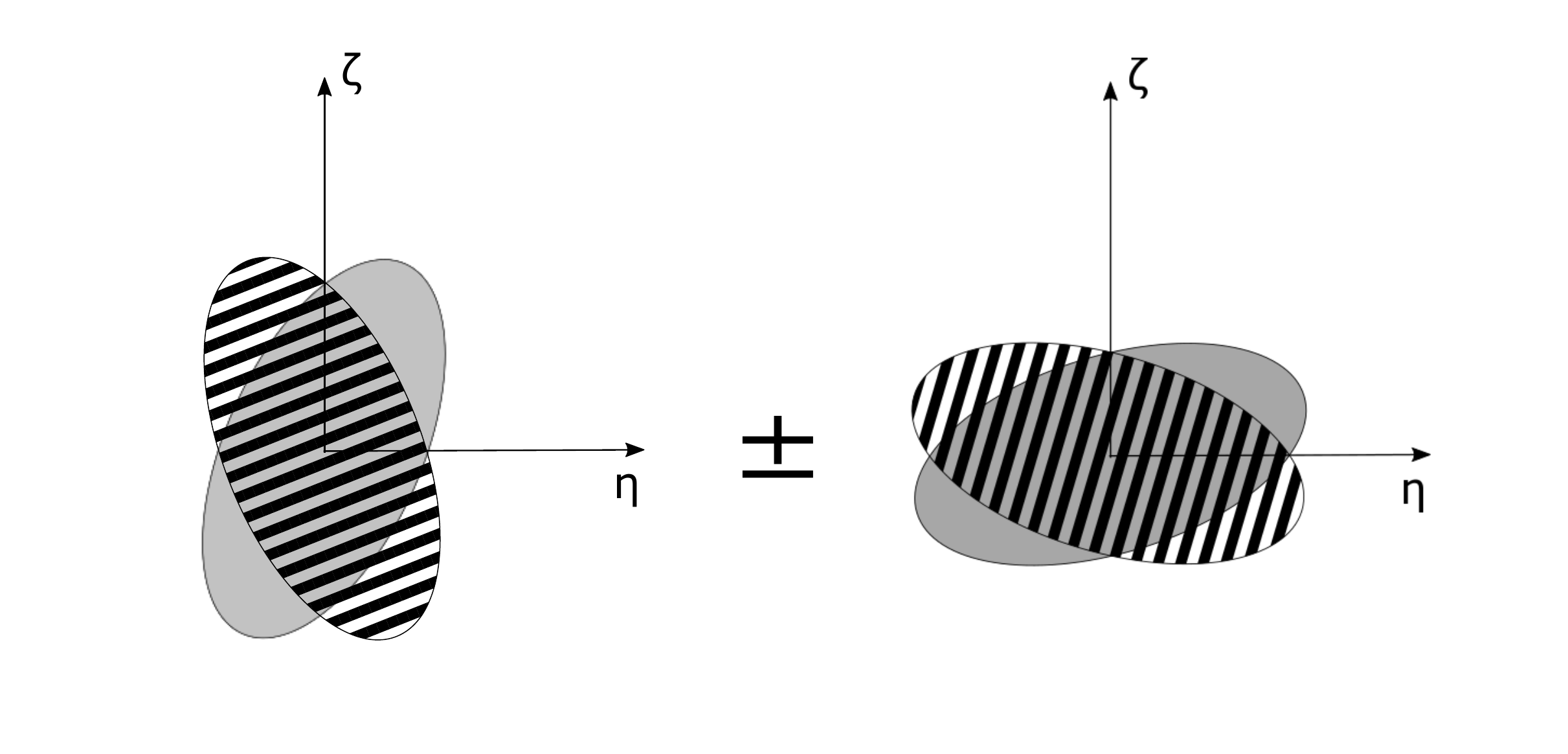}
    \end{tabular}
  \end{center}
 \caption{The TRM Hamiltonian has a double well potential, the two wells corresponding to the precession of the rotors axes about the $\zeta$- and $\eta$-axes of the intrinsic frame. The eigenfunctions  are {\it  superpositions of the states describing such precessions}. In the configuration on the left:  the precession  about the $\zeta$-axis   is one component of the scissors mode (the  centrifugal force tends to increase the angle between the rotors axes), while  the precession about the $\eta$-axis is one component of the purely rotational motion of the system as a whole (the centrifugal force tends to align the rotors axes); in the right part the precession about  the $\zeta$-axis is one component of a purely rotational motion while the precession about the $\eta$-axis is one component of the scissors mode.}
\end{figure}

The above refers to a TRM whose rotors are abstract rigid bodies. When they are made of particles, the $D_2$ symmetry is broken by the coupling of collective  to intrinsic motion, as it will be discussed in the next Section. 
 
Completing  the determination of the spectrum of the TRM regarded as a system of abstract  rigid the rotors can  have additional motivations. For instance the subsystems of magnetic nanoparticles are the macrospin which is made by electrons and therefore has intrinsic degrees of freedom and of a non magnetic  structure that is truly rigid. Free magnetic nanoparticles~[\onlinecite{Kies}] are also the system closer to atomic nuclei with respect to the systems  quoted above~[\onlinecite{Guer}]  in which one of the blades is a structure at rest. In floating magnetic nanoparticles instead  both subsystems  rotate with respect to each other.

\section{The point symmetry group of the TRM}

 Let me start from the  TRM as it results from the quantization of the classical model of two coupled rotors, with no reference to the nature of the rotors and then to the systems to which the model can possibly be applied.

The classical Hamiltonian reads~[\onlinecite{LoIu}] 
\be
H =\frac{1}{2{\mathcal I}_1}  {\vec L}_1^2 +  \frac{1}{2{\mathcal I}_2} {\vec L}_2^2 +V
\ee
where ${\vec L}_1, {\vec L}_2  $,  ${\mathcal I}_1, {\mathcal I}_2$  are the angular momenta and moments of inertia of  the rotors  and $V$  the  potential interaction between them.  If one denotes by  ${\hat \zeta}_1, {\hat \zeta}_2 $ the  unit vectors in the direction of the rotor axes, assuming the potential  to be only  a function of  the angle $2 \theta$ between them 
\be
\cos(2\theta)= {\hat \zeta}_1 \cdot {\hat \zeta}_2
\ee
 the intrinsic Hamiltonian depends only on these variables.   
Quantization of the Hamiltonian is simply  obtained by replacing the classical angular momenta by the corresponding operators. But appropriate prescriptions are necessary for the wave functions. Firstly for rigid bodies with axial symmetry rotations about the symmetry axes are not observable. Therefore if the rotors are assumed to have axial symmetry, one has to impose the constraints on the nuclear wave function
\be
{\vec L}_1 \cdot {\hat \zeta}_1 \Psi = 0, \,\,\, {\vec L}_2 \cdot {\hat \zeta}_2 \Psi = 0 \,. \label{constraints}
\ee
Secondly if the rotors have the further symmetry that inversion of their axes is not observable one has to impose the additional constraint
\begin{eqnarray} 
|\Psi({\hat \zeta}_1, {\hat \zeta}_2)|^2 =  |\Psi( r_1 \, {\hat \zeta}_1, r_2 \, {\hat \zeta}_2)|^2
\end{eqnarray}
where $r_1, r_2 = \pm1$. So the physical states must be eigenstates of the  inversion operators ${\mathcal I}_{\zeta_1}, {\mathcal I}_{\zeta_2} $ that have eigenvalues $\pm1$
\be
{\mathcal I}_{\zeta_1}\Psi = r_1 \Psi\,, \,\,\,  {\mathcal I}_{\zeta_2}\Psi= r_2 \Psi\,.
\ee
The inversion operators together with parity
\be
{\mathcal P} = {\mathcal I}_{\zeta_1} {\mathcal I}_{\zeta_2}
\ee
and the identity  constitute the point group of the TRM.
  I will classify the states  according to the eigenvalues $\pi$ of ${\mathcal P}$ and  $r_1$ of ${\mathcal I}_{\zeta_1}$, so that    for each intrinsic energy one  has the quartet $(\pi, r_1) = (++), (+-), (-+), (--)$ corresponding to $(r_2, r_1)= (++), (--), (-+), (+-)$ . 

In an actual physical system collective and intrinsic motion are coupled. From a theoretical point of view such a coupling comes from the fact  that when one introduces collective variables, one must correspondingly reduce the number of single particle coordinates. Classical ways to do it are the method of canonical transformation~[\onlinecite{Vill}] and the method of redundant variables~[\onlinecite{Sche}].  These methods, however, are conceptually illuminating but   hard to apply in practice and especially in the present case, so that  it turns out much easier  to compensate for the introduction of collective variables by putting constraints on the states. 

For an axially symmetric nucleus  invariant under inversion of its symmetry axis, for instance,
one must require that the action of inversion performed on intrinsic and collective variables  give the same result~[\onlinecite{Bohr}]
\be
{\mathcal I}_{\zeta}\Psi = {\mathcal I}_{intr \,{\zeta}} \Psi 
\ee
where $ {\mathcal I}_{intr \,{\zeta}} $ is the intrinsic inversion operators. Therefore, since the physical states must be eigenstates of the intrinsic operators 
\be 
{\mathcal I}_{intr \, {\zeta}} \Psi({\hat \zeta}) = r \Psi({\hat \zeta}) \,.
\ee
Then for  $r_i=+1, -1$, as it is well known, only even, odd,  angular momenta  are permissible. There are nuclei, however, which exhibit both even and odd angular momentum states in their spectra, which means that they can live in both the $r=\pm$ intrinsic states.  Clean illustrative examples are $^{20}$Ne among light nuclei and $^{166}$Ho among rare earths~[\onlinecite{Bohr}], pp.97, 120. The  difference of intrinsic energy between the $r=\pm$ bands is about 3.5 Mev for the first and 60 Kev for the  second nucleus. In the present case, however,  the above considerations should be applied to {\it each rotor, not to the whole nucleus. Then the crucial question is whether both $r=\pm$ states  exist for each rotor and which is the relative energy difference}. If they do exist and their energy difference is sufficiently small one can neglect  it  and to a first approximation regard the $r=\pm$ states as degenerate. Treating the rotors as truly rigid one disregards the intrinsic excitations energy necessary to build states with $r=-1$. 

In conclusion in the TRM  one must have
\be 
{\mathcal I}_{intr \, {\zeta_i}} \Psi({\hat \zeta}_1, {\hat \zeta}_2) = r_i \Psi({\hat \zeta}_1, {\hat \zeta}_2) \,.
\ee
  According to the above considerations  the members $(r_2, r_1)= (+ -), (- +),(--) $ of the $D_2$ quartets bear some analogy with scissors modes $(+ +)$ built on excited nuclear levels~[\onlinecite{Krti,Adek}]. 

Actually the situation is a bit more complicated, because the TRM wave functions are entangled and the inversion operators~\reff{inversion}  act on {\it both} the components depicted in Fig.2 that {\it result to be interchanged}.

\section{The intrinsic Hamiltonian}

Use of the 4 variables ${\hat \zeta}_i$  is cumbersome. It is instead convenient replace them 
 by  the Euler angles $\alpha, \beta, \gamma$  that describe the orientation of the system of the rotors as a whole plus the variable $\theta$.  Euler angles are associated  with the direction cosines of the axes of the intrinsic  frame  
 \begin{eqnarray}
{\hat \xi} &=&\frac{{\hat \zeta}_2 \times {\hat \zeta}_1}{ 2 \sin\theta}, \, 
{\hat \eta}=\frac{{\hat \zeta}_2 - {\hat \zeta}_1}{2 \sin\theta}, \,
{\hat \zeta}=\frac{{\hat \zeta}_2 + {\hat \zeta}_1}{2 \cos \theta}
\nonumber\\
 {\hat \zeta}_1 &=& -\sin \theta \, {\hat \eta} + \cos \theta \, {\hat \zeta}\,, \,\,\, 
{\hat \zeta}_2=\sin \theta \, {\hat \eta} + \cos \theta \, {\hat \zeta}\,.
\end{eqnarray}

The correspondence $\{ {\hat \zeta}_1, {\hat \zeta}_2 \} = \{ \alpha, \beta, \gamma, \theta \}$ is one-to-one and regular for $ 0< \theta < \frac{\pi}{2} $. These variables  are not sufficient to describe the configurations of the classical system, because they do not determine the angle of each rotor about its symmetry axis, but they describe uniquely the quantized system due to the constraints \reff{constraints}.
 One  can now express all the operators  in terms of the new variables.  To this end one  defines  the operators
 \be
{\vec I}={\vec L}_1 + {\vec L}_2, \,\,\, 
{\vec {\mathcal L}}= {\vec L}_1 - {\vec L}_2 \,.
\ee
${\vec I}$ is the total orbital angular momentum acting on the Euler angles, while ${\vec {\mathcal L}}$ is not an angular momentum, and has the  representation~[\onlinecite{LoIu}]
\be
{\vec {\mathcal L}}_{\xi}= i \frac{\partial}{\partial \theta}, \,\,\,  {\vec {\mathcal L}}_{\eta}= -\cot \theta {\vec L}_{\zeta}, \,\,\,
 {\vec {\mathcal L}}_{\zeta}=-\tan \theta  {\vec L}_{\eta} \,.
\ee 
The constraints \reff{constraints}  are satisfied by the above operators.
The  inversion operators in terms of the new variables are
\begin{eqnarray}
{\mathcal I}_{\zeta_1}&=&R_{\zeta}(\pi) R_{\xi}(\frac{\pi}{2}) R_{\theta}
\nonumber\\
 {\mathcal I}_{\zeta_2}&=& R_{\eta}(\pi) R_{\xi}(\frac{\pi}{2}) R_{\theta}
 \nonumber\\
{\mathcal P}&=& {\mathcal I}_{\zeta_1} {\mathcal I}_{\zeta_2}={\mathcal R}_{\xi}(\pi)  \label{inversion}
\end{eqnarray}
where $ R_{\zeta}(\pi), R_{\eta}(\pi),  R_{\xi}(\frac{\pi}{2})$ are rotation operators about  the intrinsic axes and
\be
R_{\theta} f(\theta)= f(\pi/2 - \theta)= \giro{f(\theta)}\,.
\ee 
The above equation shows that the whole range of $ 0 < \theta < \pi/2$ is necessary to cover the entire configurations space of the TR system.

 The transformed  Hamiltonian  is the sum of the rotational Hamiltonian of the two-rotors system as a whole plus an intrinsic Hamiltonian
\be
H= \frac{{\vec I}^2 }{2{\mathcal I}}+ H_{intr} 
\ee
where
$
{\mathcal I}= {\mathcal I}_1 {\mathcal I}_2/
({\mathcal I}_1 + {\mathcal I}_2)\,.
$
The intrinsic Hamiltonian reads
\begin{eqnarray}
H_{intr}  = \frac{1}{2{\mathcal I}}\left[ \cot^2\theta I_{\zeta}^2 + \tan \theta^2 I_{\eta}^2 - \frac{\partial^2}{\partial \theta^2}
-2 \cot(2\theta) \frac{\partial}{\partial \theta}\right] &&
\nonumber\\
+ \frac {{\mathcal I}_1 - {\mathcal I}_2}{4 {\mathcal I}_1 {\mathcal I}_2} 
\left[ -  \tan \theta I_{\zeta} I_{\eta} -  \cot\theta I_{\eta}  I_{\zeta}      
+ i  I_{\xi} \frac{\partial}{\partial \theta} \right] +V \,.&& \label{intrinsic}
\end{eqnarray}
Neglecting the term proportional to the difference of the moments of inertia in Eq.~\reff{intrinsic} and  eliminating  the linear derivative  by a unitary   transformation~[\onlinecite{DeFr}] one gets
 \begin{eqnarray}
H_{intr}' &=& U H_I U^{-1}=  { 1 \over 2  {\mathcal I} } \Bigg[ - { d^2 \over d \theta^2 } -\left( 2+  \cot^2 ( 2\theta)\right)
\nonumber\\
&+& \cot^2 \theta \, I_{\zeta}^2 + \tan^2 \theta  I_{\eta}^2  \Bigg] +V(\theta)\,. \label{Htransf}
\end{eqnarray}
At last  one assumes that the angle $\theta$   varies in such a small region  that one can perform  the harmonic approximation for the circular functions and assume a quadratic approximation for the potential
\be
V  \approx {1\over 2} C \, \theta_0^2 \Big(  x^2 s_I + (\giro{x})^2 s_{II} \Big)\,.
\ee
In the above equation
\begin{eqnarray}
&&\theta_0= \frac{\hbar}{\sqrt{ {\mathcal I} C}}\,, \,\,\, x= \frac{\theta} { \theta_0}\,,  \,\,\, 
\nonumber\\
&& s_I= s(\theta) s\Big( \frac{\pi}{4}- \theta \Big ), \,\, s_{II}= \giro{s_I}
\end{eqnarray}
where $C$ is a restoring force constant and $s(x)=1, x>0$ and zero otherwise. Such an approximation is justified by the phenomenological value $\theta_0^2\sim 0.01$ in the rare earth region. It makes  more evident   that \reff{Htransf} is  a double well Hamiltonian, implying that  {\it the rotor axes oscillate about both the $\zeta$- and $\eta$-axes}. 
 
 In the presence of a double well the eigenstates occur in doublets, whose  energy splitting can be estimated with  the WKB approximation
\be
\delta E \approx E \exp \left\{ - \int_{- \theta(E)}^{\theta(E)} d \theta \, |p(\theta)| \right\}
\ee
where $\theta(E)$  is the angle of inversion of the classical trajectory of energy $E$ and $p(\theta)$ its conjugate  momentum, 
$|p|=\sqrt{ |2 {\mathcal I}(E-V) |}$.

The   actual value  of the energy splitting in each doublet depends crucially on the potential barrier, and  I do not know the actual form of the potential beyond the harmonic approximation. I can however determine an upper bound by the (unrealistic) assumption  that the potential does not grow beyond the value $\frac{1}{2} C \theta_0^2$. Because $\theta(E) \approx \theta_0$ for the lowest states 
\be
\int_{ \theta(E)}^{\frac{\pi}{2}-\theta(E)} d \theta (-|p(\theta)|) <\frac{\pi}{2 \, \theta_0^2}
\ee
so that the energy splitting 
\be
\delta \,  E  <<  \exp \left( - \frac{\pi}{2 \, \theta_0^2} \right)  E \sim  \exp \left( - 100 \right)  E \label{split}
\ee
for typical values of $\theta_0^2 \sim 0.01$.
Such energy splitting obviously cannot be either observed or reproduced in numerical calculations and  can  therefore be altogether ignored as it was done since the beginning~[\onlinecite{LoIu}].

 I impose the normalization
\be
\int_0^{2\pi} d\alpha \int_0^{\pi}d \beta \sin \beta \int_0^{2 \pi }d\gamma \int_0^{{\pi \over 2}}d \theta \, |\Psi_{I^{\pi}M n r_1}|^2 =1
\ee
where $I,M$ are the nucleus total angular momentum and its component on the $z$-axis of the laboratory frame and $n$ labels the energy levels. I remind that  $\pi$ is the parity and $r_1$ the eigenvalue of $I_{\zeta_1} $. {\it The component $K$ of the total angular momentum on the $\zeta$-axis in general is not a good quantum number because axial symmetry is broken by the relative oscillations of the rotors}.

In the present work I study only the lowest states with $I=0,1$. 
The  eigenfunctions and eigenvalues of $H_{intr}'$ in region I are then~[\onlinecite{DeFr}]
\begin{eqnarray}
\varphi_K(x) &=& \sqrt{1\over  \theta_0} \, x^{K+{1 \over2}} \, e^{-{ 1\over 2} x^2}
\nonumber\\
{\mathcal E}_K &=& \hbar \omega (K +1)\,, \,\,\omega = \sqrt{C \over {\mathcal I}}
\end{eqnarray}
with normalization
\be
\int_0^{\infty} dx \, \left(\varphi_K(x)\right)^2 = { 1\over 2}\,.
\ee

\section{ Positive parity states}

 Consider the states 
 \be
\Psi_{I^+M n r_1} =\delta_{n,K+1} \, {\mathcal F}^I_{MK}(\alpha, \beta, \gamma) \Phi_{I^+n r_1}(\theta), \,\,\, n=1,2 \label{positive}
\ee
where
\be
{\mathcal F}^I_{MK}= \sqrt{{2I+1}\over 16( 1 +\delta_{K0}) \pi^2 } \left( {\mathcal D}^I_{MK} +(-1)^I {\mathcal D}^J_{M-K}   \right). 
\ee
 Because for the quantum numbers I=K=0; I=K=1
\begin{eqnarray}
I_{\zeta}^2 {\mathcal F}_{M,1}^1= I_{\eta}^2 {\mathcal F}_{M,1}^1 = {\mathcal F}_{M,1}^1\,, \,\,\,\,
I_{\zeta}^2 {\mathcal F}_{0,0}^0=I_{\eta}^2 {\mathcal F}_{0,0}^0 = 0 \,,
\end{eqnarray}
 the $\Phi$'s satisfy the eigenvalue equation
\begin{eqnarray}
&&\Big\{ \frac{1}{2{\mathcal I}}
\left[- \frac{\partial^2}{\partial \theta^2} + K^2(\cot^2\theta  + \tan \theta^2 ) 
  -(2+\cot^2(2\theta) )\right] 
\nonumber\\
&&+V - {\mathcal E}_{I^+Kr_1 }\Big\} \Phi_{I^+Kr_1}(\theta)=0  \label{forI=o}
\end{eqnarray}
that is symmetric under the reflection $R_{\theta}$. I then find
\begin{eqnarray}
 \Phi_{0^+, \, 1, \, r_1} &= & s_I \varphi_0 + r_1 \, s_{II} \giro{\varphi}_0 \,, \,\,\, {\mathcal E}_{0^+, \, 1, \, r_1} = \hbar \omega
\nonumber\\
 \Phi_{1^+, \, 2, \, r_1 }&= & s_I\varphi_1 - r_1 \, s_{II} \giro{\varphi}_1\,, \,\,\,  {\mathcal E}_{1^+, \, 2, \, r_1 }=2 \hbar \omega\,. 
 \label{entanglementpositive}
 \end{eqnarray}
For each  energy there is  a  doublet 
\begin{eqnarray}
\Psi_{I^+M,1,r_1} &=&{\mathcal F}^I_{M0}(s_I \varphi_0 + r_1 \, s_{II} \giro{\varphi}_0 )\,, \,\,\, {\mathcal E}_{0^+, \, 1, \, r_1} = \hbar \omega
\nonumber\\
\Psi_{I^+M,2,r_1} &=&{\mathcal F}^I_{M1}(s_I\varphi_1 - r_1 \, s_{II} \giro{\varphi}_1)\,, \,\,\,  {\mathcal E}_{1^+, \, 2, \, r_1 }=2 \hbar \omega 
\nonumber\\
\end{eqnarray}
whose members are distinguished by the intrinsic quantum number $r_1=\pm1$ 
\be
{\mathcal I}_{\zeta_1} \Psi_{I^+M n r_1}= {\mathcal I}_{\zeta_2} \Psi_{I^+M n r_1} =   r_1  \Psi_{I^+M n r_1}\,.
\label{positive}
\ee
Therefore according to \reff{inversion} all the above  states have positive parity.
 I notice that {\it the entanglement defined by \reff{entanglementpositive} is a consequence of the requirement of invariance under separate inversion of the rotors axes}. 

The states $\Psi_{0^+, 0, 1, + }$ and $\Psi_{1^+, M, 2, +}$ are the ground state and the scissors mode determined in Ref.[\onlinecite{LoIu}].

 The  strengths of magnetic dipole transitions  between states of positive parity  are
\begin{eqnarray}
B(M1; 0^+, r_1 \rightarrow 1^+, r_1 ) &=&  B(M1)\uparrow_{scissors}
\nonumber\\
B(M1;  0^+, r_1  \rightarrow 1^+, - r_1) &=&0 \,.
\end{eqnarray}
Notice that they do not vanish only between states with the same value of $r_1$ and since parity is the same, with the same value of $r_2$. All the electric dipole transition amplitudes  vanish because  relative rotations of rigid bodies can generate a magnetic dipole moment but not an electric one.

\section{Negative parity states}

Firstly I notice that  for $I=0$ there is the  unique intrinsic Hamiltonian appearing in Eq.\reff{forI=o} with $K=0$. Therefore for $I=0$ there exists only the state with positive parity determined above.  Next consider the states
 \be
\Lambda_{1^-M K n} = G^1_{MK}(\alpha, \beta, \gamma) \chi_{1^- K \, n}(\theta) \label{negative}
\ee
where
\begin{eqnarray}
G^1_{M 1} &=&  \sqrt{{3}\over 16 \pi^2 } \left( {\mathcal D}^1_{M1} + {\mathcal D}^1_{M-1}   \right)\
\nonumber\\
G^1_{M 0} &=&  \sqrt{{3}\over 16 \pi^2 } \, D_{M0}^1\,.
 \end{eqnarray}
Because
 \begin{eqnarray}
I_{\zeta}^2 G_{M,1}^1&=& G_{M,1}^1 \,,  \,\,\, I_{\eta}^2 G_{M,1}^1=0
\nonumber\\
I_{\zeta}^2 G_{M,0}^1&=&0 \,, \,\,\,
I_{\eta}^2 G_{M,0}^1 =   G_{M,0}^1
\end{eqnarray}
 the $\chi$'s satisfy the eigenvalue equations
\begin{eqnarray}
&&\Big\{ \frac{1}{2{\mathcal I}}\left[- \frac{\partial^2}{\partial \theta^2} + \cot^2\theta  
  -(2+\cot^2(2\theta) )\right] 
 \nonumber\\
&&+V -  {\mathcal E}_{1^- 1 \, \, n }\Big\} \chi_{1^- 1 \,\, n}(\theta)=0,   \,\,\,\,\,\, \,\,\,\,\mbox{for}\,\, K=1
\\
&&\Big\{ 
\frac{1}{2{\mathcal I}}\left[- \frac{\partial^2}{\partial \theta^2} + \tan^2\theta  
  -(2+\cot^2(2\theta) )\right] 
\nonumber\\
&& +V - {\mathcal E}_{1^- 0 \, \, n} \Big\} \chi_{I^-  0   \,\, n }(\theta)=0 ,
\,\,\,\,\,\, \,\,\,\,\, \mbox{for} \,\, K=0 \,.
\end{eqnarray}
These equations are not separately invariant under the reflection $R_{\theta}$, but are changed into each other.  Their solutions are
\begin{eqnarray}
\chi_{1^-,\,0,1} &= & s_I {\sqrt 2} \, \varphi_0 \,, \,\,\,\,  \chi_{1^-,\,1,1} = s_{II} {\sqrt 2} \giro{\varphi}_0 
\nonumber\\
 {\mathcal E}_{1^-, \, 0, \, 1} &=& {\mathcal E}_{1^-, \, 1, \, 1}  =\hbar  \omega
\end{eqnarray}
\begin{eqnarray}
\chi_{1^-, \,0,2 } &= &s_{II} {\sqrt 2} \giro{\varphi}_1\,, \,\,\,\,  \chi_{1^-, 1,2} = s_I {\sqrt 2} \, \varphi_1\,, \,\,\,\,\,
 \nonumber\\
 {\mathcal E}_{1^-, \, 0, \, 2} &=&  {\mathcal E}_{1^-, \, 1, \, 2}  = 2 \hbar \omega\,.
\end{eqnarray}
There are 2 degenerate states for each energy eigenvalue, distinguished by the $K$-quantum number. The above results have a simple interpretation. The state $\Lambda _{1^-M 1 1} $,   for instance,  corresponds to the configuration in the right  part of Fig.2, in which the rotation  of the system as a whole takes place about the $\zeta$-axis  without relative precession (apart that of  the zero point).  Similarly the state $\Lambda_{1^-M 0 1} $ corresponds to the configuration in the left  part of Fig.2, in which the rotation of the system as a whole  takes place about the $\eta$-axis. These two states describe rotations of the two-rotors system in its ground intrinsic state, and their energies are purely rotational. The states $\Lambda_{1^-M 1 2 }$ and  $\Lambda_{1^- M 0 2 }$ instead correspond to the right/left  part of Fig.2, but with a precession   about the $\eta, \zeta$-axis respectively. So in both cases there is  an intrinsic excitation that corresponds to the scissors mode. 

The action of a separate inversion on the rotors axes on the above wave functions is 
\begin{eqnarray}
{\mathcal I}_{\zeta_1} \Lambda_{1^-M 1 \, n} =  i \, \Lambda_{1^-M 0 \, n},\,\,\,\,
{\mathcal I}_{\zeta_1} \Lambda_{1^-M 0 \,n} = -\, i \Lambda_{1^-M 1 \,n} 
\nonumber\\
{\mathcal I}_{\zeta_2} \Lambda_{1^-M 1 \, n} = - i \Lambda_{1^-M 0\,n} , \,\,\,\,
{\mathcal I}_{\zeta_2} \Lambda_{1^-M 0 \,n}=  i  \, \Lambda_{1^-M 1 \,n}.
\end{eqnarray}
It does not merely inverts an axis, but {\it it also interchanges with each other the wave functions depicted in Fig.2. }, which therefore are not eigenstates of parity and inversions. 
 Simultaneous eigenstates of parity and inversion  are {\it superpositions of states with different $K$-quantum number, showing the breaking of axial symmetry}
 \be
\Psi_{1^- M n \,  r_1} = \frac{1}{\sqrt 2} \left(  \Lambda_{1^-M 1 n} + i r_1 \Lambda_{1^-M 0 n}
 \right)
\ee
or more explicitly 
\begin{eqnarray}
\Psi_{1^- M, 1,  \,  r_1} &=&{\mathcal G}^1_{M1} s_{II} \giro{\varphi}_0 + i r_1  {\mathcal G}^1_{M0} s_I \varphi_0
\nonumber\\
\Psi_{1^- M, 2, \,  r_1}&=&{\mathcal G}^1_{M1} s_{I} \varphi_1+ i r_1 {\mathcal G}^1_{M0} s_{II}\giro{\varphi}_1\,.
\end{eqnarray}
Because
\begin{eqnarray}
&&{\mathcal I}_{\zeta_1} \Psi_{1^- M n \, r_1} =  r_1 \, \Psi_{1^- M n \,  r_1} 
\nonumber\\
&&{\mathcal I}_{\zeta_2} \Psi_{1^- M n \,  r_1} = - r_1 \, \Psi_{1^- M n \,  r_1}\,. \label{negative}
\end{eqnarray}
these states have negative parity.
  So  there are   2 doublets of negative parity states of energy $\hbar  \omega, 2 \hbar  \omega$ respectively, and {\it again we see that   the entanglement is originated by the requirement of invariance under  inversions of the rotor axes}.  
 
The  strengths of magnetic dipole transitions  between states of negative  parity  are
\begin{eqnarray}
 B\left(M1; ( 1^-, \, 2, \, r_1)  \rightarrow    (1^-, \, 1,  r_1 ) \right) &=& \frac{1}{2} B\downarrow_{scissors} 
 \nonumber\\
 B\left(M1; ( 1^-, \, 2, \, r_1)  \rightarrow    (1^-, \, 1,  - r_1 ) \right) &=& 0 \,. \label{strenghtnegpar}
\end{eqnarray}
They connect only states with the same value of the  $r_1$ quantum number and since parity is the same, with the same value of $r_2$. There are no electromagnetic transitions between the states of positive and negative parity because the only operator which could connect them, the magnetic quadrupole operator, evaluated according to the by now standard procedure~[\onlinecite{LoIu, Palu}] vanishes identically. {\it The lowest negative parity scissors mode is therefore stable within the TRM}.

\section{The $D_2$ quadruplets}

 In summary the  two lowest $D_2$  quadruplets of the TRM are
 \begin{eqnarray}
\Psi_{0^+ 0 1 \,\pm}  \,,   \, \Psi_{1^- M 1 \, \pm}\,, \,\,\, &&  {\mathcal E}_1= \hbar \omega
\nonumber\\
\Psi_{1^+ M 2 \,\pm}  \,,  \Psi_{1^- M 2 \, \pm}\,, \,\,\, &&{\mathcal E}_2 = 2 \hbar \omega\,.
\end{eqnarray}
The quadruplet  of  intrinsic  energy $\omega$ contains the  states 
$\Psi_{0^+ 0 1, r_1 }$ of positive parity and the purely rotational states $ \Psi_{1^- M 1, r_1 } $ of negative parity. Their total energies  differ by the purely  rotational energy $3  \hbar^2 / (2 {\mathcal I}) $.  The quadruplet of intrinsic  energy $2 \hbar \omega$ contains the positive parity state $ \Psi_{1^+ M 2 ,+} $, the known scissors mode, plus the positive parity state $ \Psi_{1^+ M 2 , -} $ and the negative parity states  $ \Psi_{1^- M 2 \, \pm}$.

In the application of the TRM  to atomic nuclei, however,  one must take into account the coupling  between collective and intrinsic variables that I discussed in Section 2. The degeneracy related to the $r_1$-quantum number within each multiplet should  then be broken: states with  negative $r$ should have an intrinsic  energy  higher  than states with positive $r$. If I call $e$ the energy necessary to excite a $r=-1$ state of a rotor, the states of the first quartet $\Psi_{0^+ 0 1, + },  \Psi_{1^- M 1, \pm }, \Psi_{0^+ 0 1, - } $ have energies $\hbar \omega, \hbar \omega +e, \hbar \omega+2e$ respectively, and the states $ \Psi_{1^+ M 2 ,+},  \Psi_{1^- M 2 \, \pm},  \Psi_{1^+ M 2 , -} $ of the second quartet have energies $2\hbar \omega, 2\hbar \omega +e, 2\hbar \omega+2e$ respectively. Notice however, that $M1$ transitions occur only between states in which  the rotors are in the same state, and then their energies  differ by $\hbar \omega$. Of course there will also be transitions of positive or negative parity in which the rotors change the $r$-quantum number, but I cannot say anything about them.

\section{ Conclusions}

 The intrinsic Hamiltonian of the TRM with rigid rotors has a $D_2$ point symmetry. Because of it the eigenstates occur in quadruplets that contain degenerate positive and negative parity states. In actual nuclei, if all these states really occur,  such a degeneracy  should be  broken by the coupling between collective and intrinsic variables,  states in which one or both rotors have   negative $r$ having an intrinsic  energy  $\omega+e, \omega+2e$ respectively. 
 
 The existence of the whole quadruplets depends on 2 crucial conditions
 
 1) The nuclear states should be entangled. This property can be tested by measuring the $B(M3)$ strength in the scissors rotational band~[\onlinecite{Palu}], a measurement that appears to be feasible.  The theoretical investigation of entanglement by means of direct microscopic  calculations appears to me instead  difficult. I think that one way to do it i is to  derive from microscopic Hamiltonians  a collective Hamiltonian restricted to all the states with the $D_2$ quantum numbers determined above, following the procedures developed for the positive parity states~[\onlinecite{Diep}]. The different possibilities that arise have been discussed in~[\onlinecite{Palu}].

2) The $r=-1$ excited states of the proton and neutron fluids should exist and be sufficiently stable to support scissors modes. 
 This is what I see as a major question mark.
 
 A  comment is in order concerning the actual existence of entanglement in terrestrial atomic nuclei. This depends on how and when they were created. If they were created   with their wave functions localized in one of the potential wells, for instance, one should   know the time necessary for them to tunnel to the entangled configuration, and this would in turn require to know the  form of the potential barrier (that I assumed  to be flat for the upper  bound \reff{split}). 

  Some hint on the existence of the new states  might perhaps be obtained  by  two step cascade experiments~[\onlinecite{Krti,Adek}], in which  one should observe the contribution $  B\left(M1; ( 1^-, \, 2, )  \rightarrow    (1^-, \, 1 ) \right) = \frac{1}{2} B\downarrow_{scissors} $ to the gamma strength  while the decay of the final state should have a small strength.

\end{document}